\documentstyle[aps,pra,epsfig,multicol]{revtex}

\newcommand{\ket}[1]{\mbox{$|#1\rangle$}}

\newcommand{\op}[1]{\mbox{\boldmath $\hat{#1}$}}

\begin{document}

\title{Signatures of the Pair-Coherent State}
\author{A. Gilchrist$^{*}$ and W.J. Munro$^{\dagger}$}
\address{$^{*}$ School of Computing and Mathematical Sciences, University of
Waikato, Hamilton, New Zealand}
\address{$^{\dagger}$ Centre for Laser Science,Department of Physics, University of
Queensland,\\ QLD 4072, Brisbane, Australia}

\date{\today}

\maketitle

\begin{abstract}
  We explore in detail the possibility of generating a pair-coherent
  state in the non-degenerate parametric oscillator when decoherence is
  included.  Such states are predicted in the transient regime in
  parametric oscillation where the pump mode is adiabatically eliminated.
  Two specific signatures are examined to indicate whether the state of
  interest has been generated, the Schr\"odinger cat state---like
  signatures, and the fidelity. Solutions in a transient regime reveal
  interference fringes which are indicative of the formation of a
  Schr\"odinger cat state. The fidelity indicates the purity of our
  prepared state compared to the ideal pair-coherent state.
\end{abstract}

\begin{multicols}{2}
        
\section{Introduction}

Quantum mechanics is well known for two fundamental quantities, the
superposition of states and entanglement. Both have been studied
extensively and much rich physics has been observed. Generally however
systems exhibit one but not generally both of these phenomena.  Recently
however the idealised state produced in non-degenerate parametric
oscillation when the pump mode is adiabatically eliminated has been shown
to give the possibility of observing both Schr\"odinger cat states
\cite{RK93-552} and producing correlated photon pairs which could be used
to perform tests of quantum mechanics versus local realism such as the EPR
paradox\cite{TA94-2870} and the Bell inequality\cite{GDR98-3169,GDR98-4259}.

There is still much interest in the possibility of generating
experimentally a Schr\"odinger cat state. Such a state is defined as a
quantum superposition of two macroscopically distinct states
\cite{S35-844}. It was Schr\"odinger's concern that quantum mechanics does
not fundamentally prohibit the existence of such states, which seemingly
defy physical reality.  Work by Krippner and Reid \cite{RK93-552} predicts
that a Schr\"odinger cat state may be produced in the signal field of
non-degenerate parametric oscillation, in a transient regime.  The
theoretical model presented here incorporates the effect of linear signal
losses, which tend to oppose the formation of the superposition state.

Recent work by Gilchrist {\it et. al.} \cite{GDR98-3169,GDR98-4259} and Munro
\cite{99m4197} showed how in a certain narrow regime, the pair-coherent
state gives quantum mechanical predictions that are in disagreement with
those of local hidden variable theories for a situation involving
continuous quadrature phase amplitude measurements.  This test could be
achieved by binning the continuous position and momentum information into
two categories and using the binary results in the strong Clauser Horne
Bell inequality test\cite{CH74-526}. The predicted violation was small
(less than 2\%).  Such results were highly idealised, and assumed the
preparation of a pair-coherent state to begin with.

In this paper we will investigate the possibility of generating a
pair-coherent state in the non-degenerate parametric oscillator with an
adiabatically eliminated pump mode when decoherence is included.  Two
specific signatures of the state will be considered, interference fringes
which are indicative of the formation of a Schr\"odinger cat state and the
fidelity. In particular the fidelity will provide an indication of the
parameter regime required to perform a loophole free test of quantum
mechanics.

It should be noted that the fidelity is a rather abstract quantity and in
practice would require the entire state to be reconstructed in order to be
measured. The Schr\"odinger cat state signatures however are more
operationally accessible though given the sensitivity of Schr\"odinger cat
state itself to decoherence \cite{BVK92-6570} we would expect similar sensitivity
for the pair-coherent state and this is indeed the case.

\section{The Pair-Coherent State}

The following two-mode entangled quantum superposition state,
\begin{equation}
  \ket {\mbox{circle}}_{m} = {\mathcal{N}} \int_{0}^{2\pi}    
e^{-im\varsigma}
  |r_{0}e^{i\varsigma}\rangle_{a}
  |r_{0}e^{-i\varsigma}\rangle_{b} d\varsigma 
  \label{eqn:circlestate}
\end{equation}
is known as the was pair-coherent (or ``circle'') state and was
originally discussed by Agarwal \cite{A86-827,A88-1940} and Reid and
Krippner\cite{RK93-552}.  In equation~(\ref{eqn:circlestate})
$|{\ldots}\rangle_{a}$ and $|{\ldots}\rangle_{b}$ represent coherent states
in the modes $\hat a$ and $\hat b$, where the $\hat a$ and $\hat b$ are the
usual boson operators.  ${\mathcal{N}}$ is a normalisation coefficient and
$r_0$ the amplitude of the coherent state. $m$ is the photon number
difference between the signal and idler modes.  This state is actually a
continuous superposition of coherent states in a circle (hence the notation
$|\mbox{circle}\rangle$). For our purposes in this paper we shall
concentrate on the $m=0$ (equal photon number in each mode) situation in
equation~(\ref{eqn:circlestate}) which when normalised is given by
\begin{equation}
  \ket {\mbox{circle}}={\mathcal{B}}^{1/2}\int_{0}^{2\pi }d\varsigma 
\ket {r_{0}e^{i\varsigma }}_{a}
  \ket {r_{0}e^{-i\varsigma }}_{b},
\end{equation}
where 
 \begin{equation}
   {\mathcal{B}}^{-1}=4\pi^{2}e^{-2r_{0}^{2}}I_{0}(2r_{0}^{2}),
 \end{equation}
 Here $I_{0}$ is a zeroth order modified Bessel function. Such a state can
 also be written in terms of correlated photon number pairs of the form
\begin{eqnarray}\label{correlatedpair}
| \Psi \rangle = \sum_{n=0}^{\infty} c_{n} | n \rangle | n \rangle 
\end{eqnarray}
where
\begin{eqnarray}\label{circlecn}
c_{n}={{r_0^{2 n}}\over{n! I_{0}\left(2 r_0^{2} \right)}}.  
\end{eqnarray}
Such a state should not be confused with the state produced by the
non-degenerate parametric amplifier (NDPA) which can also be written in the
form of (\ref{correlatedpair}) but now with the $c_{n}$ coefficients given
by
\begin{eqnarray}\label{parampcn}
c_{n}={{\tanh^n \left[\chi \epsilon \tau\right]}\over{\cosh \left[\chi \epsilon \tau\right] }}
\end{eqnarray}
Here  $\epsilon$ represents the field amplitude of a non-depleting classical 
pump, $\chi $ is proportional to the susceptibility of the medium and $\tau$ 
is the time that the modes spend in the crystal.

In this paper we are interested in exploring the generation of the
pair-coherent state.

\section{The Non-degenerate Parametric Oscillator}

It has been suggested by Reid and Krippner that the NDPO transiently 
generates a state of the above form, in the limit of very large 
parametric nonlinearity and high-$Q $ cavities \cite{RK93-552}. 
The non-degenerate parametric oscillator with linear damping in the
signal and idler modes and  the pump mode pumped by a classical field 
may be represented by the Hamiltonian
\begin{eqnarray}
  H & = & H_{I}+H_{p}+H_{{\rm irrev}}\label{e:fullH} \\
  H_{I} & = & i\hbar \kappa (a_{3}a^{\dagger }_{2}a^{\dagger }_{1}-
  a^{\dagger }_{3}a_{2}a_{1})\nonumber \\
  H_{p} & = & i\hbar \epsilon (a^{\dagger }_{3}-a_{3})\nonumber \\
  H_{{\rm irrev}} & = & \sum_{jl}(a_{j}\Gamma_{jl}^{\dagger }+
  a_{j}^{\dagger }\Gamma_{jl}).\nonumber 
\end{eqnarray}
Here $a_i$ are boson operators for the cavity modes at frequencies
$\omega_i$, where $\omega_3 = \omega_1+\omega_2$.  The mode $a_3$ is driven
by a resonant external driving field with amplitude proportional to
$\epsilon$ and is known as the pump-mode. Modes $a_1$ and $a_2$ are the
signal- and idler-modes.  The loss of photons through the cavity mirrors is
modeled by the Hamiltonian term $H_{{\rm irrev}}$, which denotes a
coupling of the cavity modes to the zero temperature reservoir modes
(symbolised by $\Gamma_i$) external to the cavity. We will denote the
cavity decay rates for the modes $a_i$ by $\gamma_i$.

Following standard techniques it is easy to derive a master equation 
of the form 
\begin{eqnarray}
\dot{\rho } & = & \frac{1}{i\hbar }[H_{I}+H_{p},\rho ]\nonumber \\
 &  & +\sum_{j=1}^{3}\gamma_{j}(2a_{j}^{\dagger }\rho 
a_{j}-a_{j}^{\dagger }a_{j}\rho -\rho a_{j}^{\dagger 
}a_{j})\label{e:master} 
\end{eqnarray}
In the limit where the pump mode is heavily damped compared to the 
other modes ($\gamma_{3}\gg \gamma_{2},\gamma_{1}$) the pump variables can 
be eliminated. This is equivalent to studying the following model 
Hamiltonian:
\begin{eqnarray}
H & = & i \hbar \frac{\epsilon \kappa }{\gamma_{3}}(a^{\dagger }_{2}
a^{\dagger }_{1}-a_{2}a_{1})\nonumber \\
 &  & 
-\frac{\kappa^{2}}{\gamma_{3}}\sum_{l}(a_{2}a_{1}\Gamma^{\dagger 
}_{l}+a^{\dagger }_{2}a^{\dagger }_{1}\Gamma_{l})\nonumber \\
 &  & +\sum_{jl}(a_{j}\Gamma_{jl}^{\dagger }+a_{j}^{\dagger 
}\Gamma_{jl}).\label{e:modelH} 
\end{eqnarray}
The presence of the two-photon damping term is the fundamental difference
between the NDPO in this limit and the NDPA.

To aid our discussion of parameters below, we will briefly examine
realistic parameter values for the non-degenerate parametric oscillators
containing the commonly used crystals, silver gallium selinide (${\rm
  AgGaSe_{2}}$) and potassium titanyl phosphate (KTP). In Table
(\ref{table_1}) are shown some typical values.  We easily observe that the
nonlinear coupling constant is much weaker than the damping constant and
hence our scaled parameter $g^{2}=\kappa^{2} / \gamma \gamma_{3}$ that we
will introduce shortly will be very small.
\begin{table}
\begin{tabular} {ccccccc}
CRYSTAL & $\kappa(s^{-1})$ &  $ \gamma(s^{-1})$ & $\kappa/\gamma$ \\ \tableline
${\rm AgGaSe_{2}}$ & $4.4 \times 10^{4}$ & $7.5\times10^{8}$ & $5.9\times10^{-5}$\\
KTP & $7.6 \times 10^{3}$ & $7.5\times10^{8}$ & $1\times10^{-5}$  \\
\end{tabular}
\caption{Table of realistic values for the nonlinear coupling constant
$\kappa$ and the damping constant $\gamma$ for two types of parametric
crystal ${\rm AgGaSe_{2}}$ and $KTP$.}
\label{table_1}
\end{table}

\section{The adiabatically eliminated Master Equation}

The Hamiltonian (\ref{e:modelH}) above corresponds to the following master
equation (equation~(\ref{e:master}) with the pump mode adiabatically
eliminated)
\begin{eqnarray}
{{d\rho }\over{d \tau}} & = & \lambda
[a^{\dagger }_{2} a^{\dagger }_{1}-a_{2}a_{1},\rho ]\nonumber \\
&  & -g^{2} (2a_{1}^{\dagger }a_{2}^{\dagger }\rho 
a_{1}a_{2}-a_{1}^{\dagger }a_{2}^{\dagger }a_{1}a_{2}\rho -\rho a_{1}^{\dagger}
a_{2}^{\dagger}a_{1}a_{2})\nonumber \\
&  & +\sum_{j=1}^{2}(2a_{j}^{\dagger }\rho 
a_{j}-a_{j}^{\dagger }a_{j}\rho -\rho a_{j}^{\dagger 
}a_{j})\label{e:master-adiabatic} 
\end{eqnarray}
where we have introduced the following scaled variables 
$\lambda=\frac{\epsilon \kappa }{\gamma_{3}\gamma}$ and 
$g^{2}=\frac{\kappa^{2}}{\gamma_{3}\gamma}$. The time has been 
scaled such that $\tau=\gamma t$ and we have assumed that the signal and 
idler decay constant $\gamma_{1}$ and $\gamma_{2}$ are in 
fact equal to $\gamma$.

The master equation can be solved numerically by projecting the master 
equation onto an infinite number state basis \cite{94gp1792}.
Expanding the density matrix in the number state basis as follows
\begin{eqnarray}\label{super.6}
\rho_{n_{1}n_{2};m_{1}m_{2}}=\langle n_{1} |\langle n_{2} |\rho | m_{1} \rangle | m_{2} \rangle,
 \end{eqnarray}
we may express the time evolution of the system as
\begin{eqnarray}\label{super.7}
{{\partial}\over{\partial \tau}} \rho_{i_{1}i_{2};j_{1}j_{2}}&=&
\langle i_{1} |\langle i_{2} |{{\partial}\over{\partial \tau}} 
\rho | j_{1} \rangle | j_{2} \rangle \nonumber \\
&=&{\cal {L}}_{i_{1}i_{2};j_{1}j_{2}}^{n_{1}n_{2};m_{1}m_{2}}
\rho_{n_{1}n_{2};m_{1}m_{2}}
\end{eqnarray}
where this super matrix ${\cal {L}}_{i_{1}i_{2};j_{1}j_{2}}^{n_{1}n_{2};m_{1}m_{2}}$ is given by 
\begin{eqnarray}\label{super.8}
{\cal {L}}_{i_{1}i_{2};j_{1}j_{2}}^{n_{1}n_{2};m_{1}m_{2}}&=&
\lambda\sqrt{i_{1}i_{2}}\; \delta_{i_{1},j_{1}}^{n_{1}+1,m_{1}} 
\delta_{i_{2},j_{2}}^{n_{2}+1,m_{2}}\nonumber \\
&-&\lambda\sqrt{\left(i_{1}+1\right)\left(i_{2}+1\right)}\;\delta_{i_{1},j_{1}}^{n_{1}-1,m_{1}} 
\delta_{i_{2},j_{2}}^{n_{2}-1,m_{2}}\nonumber \\
&+&\lambda \sqrt{j_{1}j_{2}}\;\delta_{i_{1},j_{1}}^{n_{1},m_{1}+1} 
\delta_{i_{2},j_{2}}^{n_{2},m_{2}+1}\nonumber \\ 
&-&\lambda \sqrt{\left(j_{1}+1\right)\left(j_{2}+1\right)}\; \delta_{i_{1},j_{1}}^{n_{1},m_{1}-1} 
\delta_{i_{2},j_{2}}^{n_{2},m_{2}-1}\nonumber \\
&-&2 g^{2} \prod_{k=1}^{2}\sqrt{\left(i_{k}+1\right)\left(j_{k}+1\right)}
\;\delta_{i_{1},j_{1}}^{n_{1}-1,m_{1}-1} \delta_{i_{2},j_{2}}^{n_{2}-1,m_{2}-1}\nonumber \\
&+&g^{2}\left[i_{1}i_{2}+j_{1}j_{2}\right]\delta_{i_{1},j_{1}}^{n_{1},m_{1}} \delta_{i_{2},j_{2}}^{n_{2},m_{2}}\nonumber \\
&+&2 \sqrt{\left(i_{1}+1\right)\left(j_{1}+1\right)}\;\delta_{i_{1},j_{1}}^{n_{1}-1,m_{1}-1} \delta_{i_{2},j_{2}}^{n_{2},m_{2}}\nonumber \\
&-&\left[i_{1}+j_{1}\right]\delta_{i_{1},j_{1}}^{n_{1},m_{1}} 
\delta_{i_{2},j_{2}}^{n_{2},m_{2}}\nonumber \\
&+&2 \sqrt{\left(i_{2}+1\right)\left(j_{2}+1\right)}\;\delta_{i_{1},j_{1}}^{n_{1},m_{1}}  \delta_{i_{2},j_{2}}^{n_{2}-1,m_{2}-1} \nonumber \\
&-&\left[i_{2}+j_{2}\right]\delta_{i_{1},j_{1}}^{n_{1},m_{1}} \delta_{i_{2},j_{2}}^{n_{2},m_{2}}
\end{eqnarray}
Here  

\begin{eqnarray}
\label{super.9}
\delta_{i,j}^{n,m}=\left\{\begin{array}{ll}1 & {\mbox{if} \;i=n \;\mbox{and}\; 
j=m }\\
 0 & {\rm{ otherwise}}\end{array}\right.
\end{eqnarray}

To allow for numerical calculations, one must put a finite limit on the
number of Fock states used in the basis in (\ref{super.7}).  Care must be
taken to ensure that the truncation of the number state basis is done
correctly so the population of the higher order states is small.  In
practice we found $n_{\mbox{max}}=20$ sufficient for most of the
calculations.

\section{Entanglement}

As we mentioned in the introduction, this pair-coherent state has the
property that it contains sufficient entanglement to violate a Bell
inequality.  Gilchrist {\it et. al.} showed that the pair-coherent
state specified by (\ref{eqn:circlestate}) theoretically violated a Bell
inequality. To be more explicit, they showed how using highly efficient
quadrature phase homodyne measurements, the Clauser Horne strong Bell
inequality could be tested in an all optical regime.  While the violation
may be small the highly efficient detection means that provided the
extremely ideal state could be generated, a significant test could be done.

There are a number of measures to determine the purity of the produced 
state. The measure we will use here is the fidelity. The fidelity may 
be defined as 
\begin{equation}
F=\left| \langle \mbox{circle}|\mbox{output}\rangle 
\right|^{2}
\end{equation}
in terms of pure states. In terms of the density operator $\op{\rho} $ of 
the output state, we represent the fidelity as
\begin{equation}
F={\rm Tr} \left[\rho_{\mbox{circle}}^{1/2}\;\; \rho_{\mbox{output}} \;\;
\rho_{\mbox{circle}}^{1/2}   \right]^{1/2}
\end{equation}

\begin{figure}
\epsfig{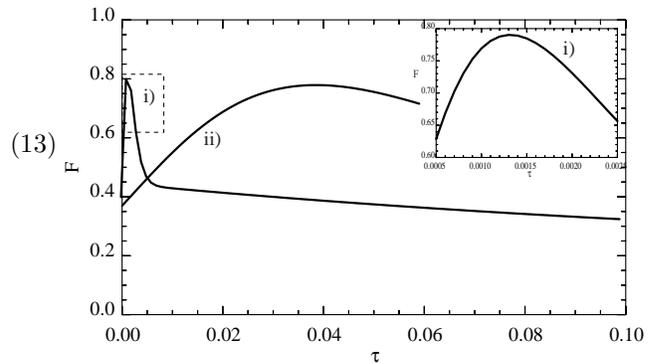}
\narrowtext
\caption{Plot of the fidelity of the generated state compared with 
  the pair-coherent state for $\lambda/g^2=1.12$ and (i) $g^2=300$ giving a
  maximum of $F\sim 0.78$, (ii) $g^2=10$ giving a maximum of $F\sim 0.76$.}
\label{fig5}
\end{figure}

Figure~\ref{fig5} shows the result of calculating the fidelity against an
ideal pair-coherent state for $\lambda/g^2=1.12$ and two different values
of $g$.  It shows a maximum fidelity of around $80\%$ for $g^2=300$. For
larger $g$ we can get a larger fidelity but the transient period over which
this is available is significantly shorter. Given the small size of the
Bell inequality violation, and the narrow parameter regime over which it
occurs in reference~\onlinecite{GDR98-3169} it is likely that higher
nonlinearities would be required for the NDPO to produce the state
sufficient to violate the Bell inequality in that scheme.

\section{Schr\"odinger Cat States}

Given a pair-coherent state, previous work by Krippner and Reid
\cite{RK93-552} has predicted that in the limit of very large $r_0$ and
with a conditional measurement of one mode, a Schr\"odinger cat state may
be produced in the other mode. Observing these states then constitutes an
indirect signature of the presence of a pair-coherent state as well as
being of interest in its own right. In this section we examine the
formation of these states when the parameters are not so extreme and with
the presence of damping.

Let us now suppose that one measures the quadrature phase amplitude 
defined by 
\begin{eqnarray}\label{super.111}
X_{\theta_{i}}&=&\left(a_{i} e^{-i \theta_{i}} +a_{i}^\dagger e^{i \theta_{i}}\right)/\sqrt{2} 
\end{eqnarray}
where $\theta_{i}$ is the phase of the local oscillator for the $i$th mode.
The two measured quadratures $X_0$ and $X_{\pi/2}$ are non commuting
observables.  We assume that the measurement of the quadrature
$X_{\theta_{i}}$ gives the result $x_{\theta_{i}}$.

It is then possible to construct the joint probability distribution 
of obtaining a result  $x_{\theta_{1}}$ for the first mode and  
$x_{\theta_{2}}$ for the second. This probability is expressed as
\begin{eqnarray}
P_{\theta_{1},\theta_{2}}\left(x_{1},x_{2}\right)=\langle x_{1} | 
\langle x_{2} |
\rho | x_{2} \rangle  | x_{1} \rangle 
\end{eqnarray}
Here we have abbreviated $x_{\theta_{1}}$ by $x_{1}$ and $x_{\theta_{2}}$
by $x_{2}$. Our state of interest can then be detected by observing
interference fringes in the probability distribution
$P_{\pi/2,0}\left(x_{1}=z,x_2=0\right)$ for the quadrature phase amplitude
measurement performed on the signal mode, conditioned on the idler mode
result $x_{2}=0$ for $\theta_{2}=0$. The observation of interference
fringes present in one of the quadrature measurements (in conjunction with
the observation of twin isolated peaks in the conjugate quadrature phase
amplitude) are indicative of Schr\"odinger cat states, where $X_0$ is
analogous to the r\^ole of position and $X_{\pi/2}$ of momentum.

In terms of our number state basis expansion for the density matrix, 
this joint probability distribution can be written as 
\begin{eqnarray}
P_{\theta_{1},\theta_{2}}\left(x_{1},x_{2}\right)= 
\sum_{{{n_{1},n_{2}}\atop{m_{1},m_{2}}}=0}^{\infty}
\rho_{n_{1}n_{2};m_{1}m_{2}} \prod_{i=1}^{2} \langle x_{\theta_{i}}|n\rangle  
\langle m|x_{\theta_{i}}\rangle 
\end{eqnarray}
where $\langle x_{\theta_{i}}|n\rangle $ is given by 
\begin{eqnarray}
\langle x_{\theta}|n\rangle&=&{{e^{-i n \theta}}\over{
\sqrt{2^n n!\sqrt{\pi}}}}\exp \left[-{1\over 2} x_{\theta}^2\right] H_n\left(x_{\theta}\right) 
\end{eqnarray}
where the units have been chosen such that $\hbar=\omega=c=1$ and $H_n\left(x_{\theta}\right)$ 
is the Hermite polynomial.

\subsection{The ideal situation}
\begin{figure}
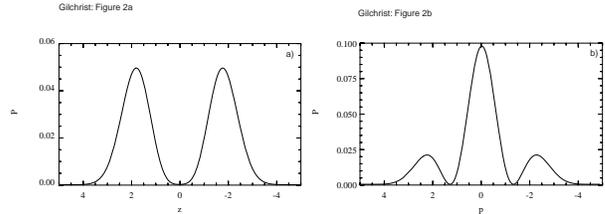

\epsfig{figure=idealprobx.epsf,width=.45\columnwidth}
\epsfig{figure=idealprobp.epsf,width=.45\columnwidth}
\narrowtext
\caption{Plot of position ($X_0$) (a) and momentum ($X_{\pi/2}$) (b) 
probability distribution for the signal mode conditioned on a 
quadrature position measurement
on the idler mode recording 0. Here we have $\lambda/g^{2}=1.5$.}
\label{fig1}
\end{figure}

In the absence of damping the ideal pair-coherent state is given by
(\ref{correlatedpair}) with the $c_{n}$ coefficients specified by
(\ref{circlecn}). In figure~\ref{fig1} we plot
$P_{0,0}\left(x_{1}=z,x_{2}=0\right)$ versus $z$ and
$P_{\pi/2,0}\left(x_{1}=p,x_{2}=0\right)$ versus $p$. We clearly observe
the interference fringes and twin peaks that characterise the Schr\"odinger
cat state $|i\lambda/g^2\rangle+|-i\lambda/g^2\rangle$.

For comparison, the Schr\"odinger cat state signatures are also shown for
the pair-coherent state required by Gilchrist {\it et. al.}
($\lambda/g^2=1.12$).  As can be expected we do not get very distinct
interference fringes and peak-resolution. Interestingly, clear cat
signatures require only slightly larger values of $\lambda/g^2$.

\begin{figure}
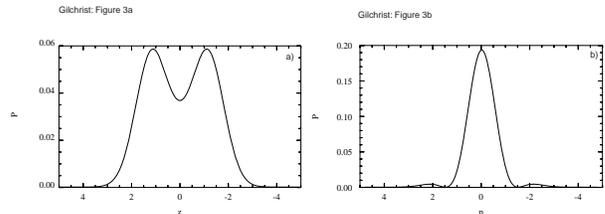

\epsfig{figure=idealprobxc.epsf,width=.45\columnwidth}
\epsfig{figure=idealprobpc.epsf,width=.45\columnwidth}
\narrowtext
\caption{Plot of position ($X_0$) (a) and momentum ($X_{\pi/2}$) (b) 
  probability distribution for the signal mode conditioned on a quadrature
  position measurement on the idler mode recording 0. Here we have
  $\lambda/g^{2}=1.12$.}
\label{fig1b}
\end{figure}

\subsection{Numerical Simulations}

The probability distribution is then written as
\begin{eqnarray}
P_{\theta_{1},\theta_{2}}\left(x_{1},x_{2}\right)= 
\sum_{{{n_{1},n_{2}}\atop{m_{1},m_{2}}}=0}^{n_{{\rm max}}}
\rho_{n_{1}n_{2};m_{1}m_{2}} \prod_{i=1}^{2} \langle x_{\theta_{i}}|n\rangle  
\langle m|x_{\theta_{i}}\rangle 
\end{eqnarray}
 In our calculations the effect on increasing the number of basis 
states by one produced an error of less than $0.001$ percent. 

\begin{figure}
\epsfig{figure=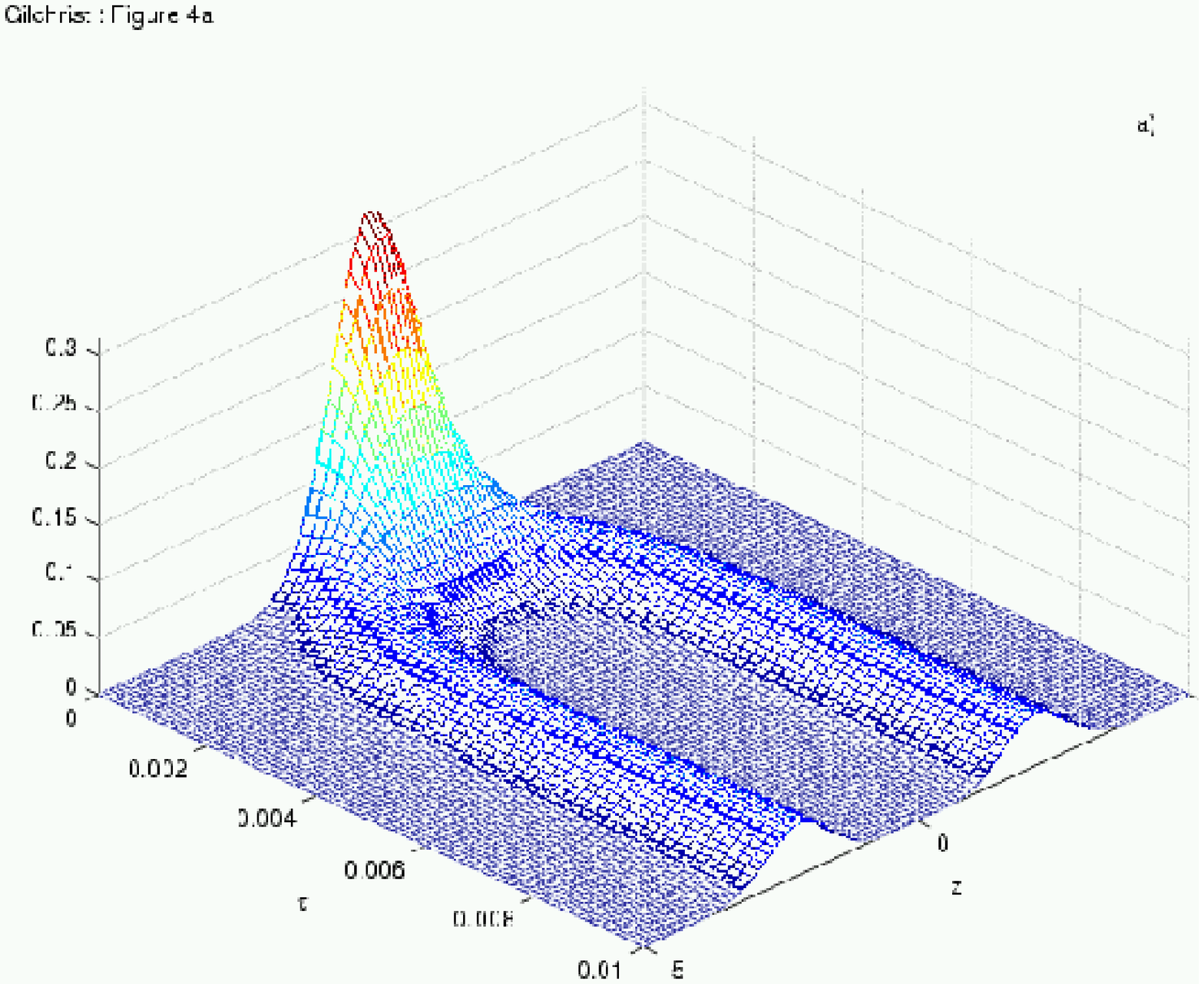,width=.9\columnwidth}
\epsfig{figure=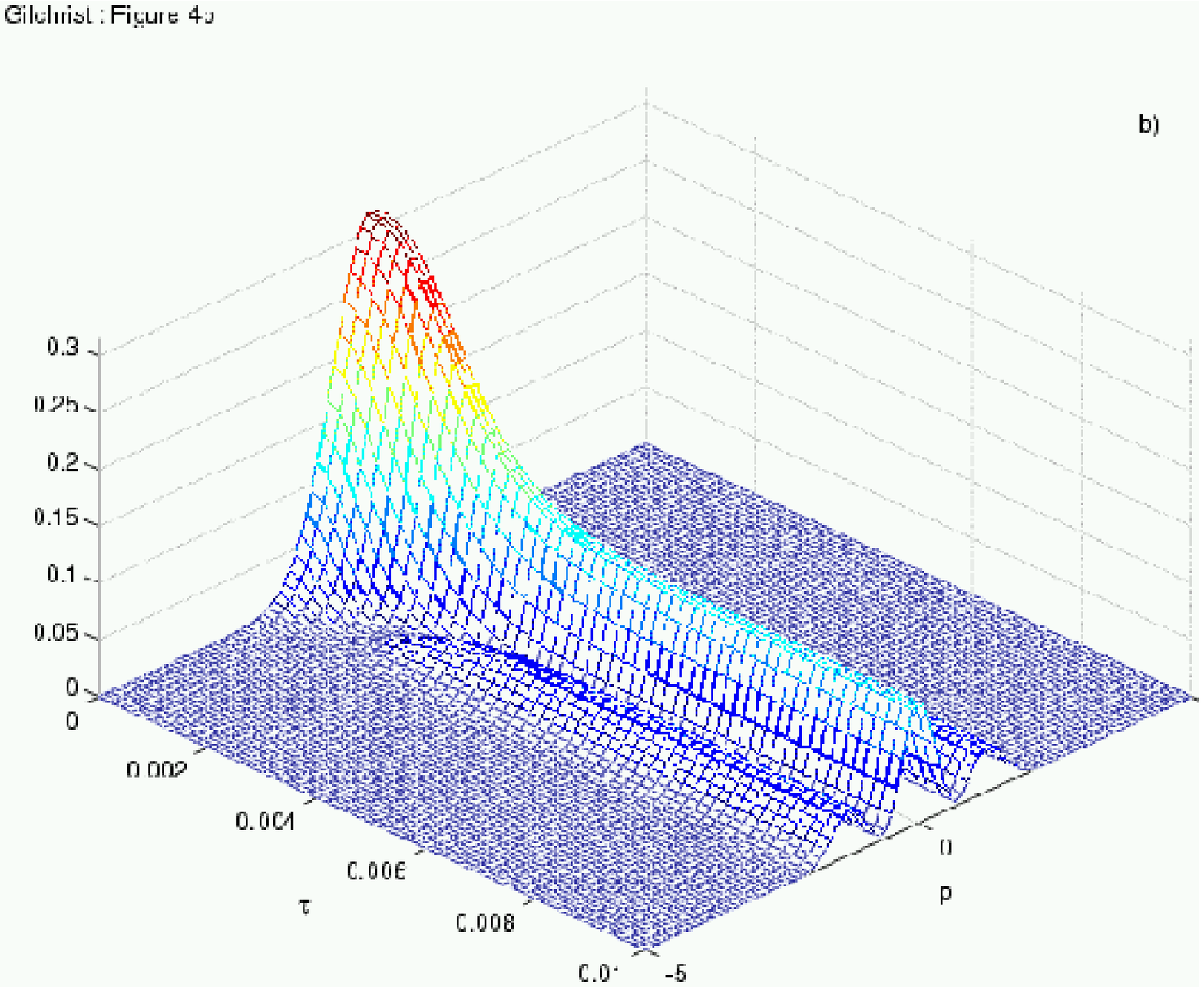,width=.9\columnwidth}
\narrowtext
\caption{Plot of (a) the evolution of the position probability 
  distribution $P(z)$, and (b) the momentum probability distribution
  $P(p)$. Here $g^{2}=300$ with $\lambda/g^2=1.5$.}
\label{fig2}
\end{figure}

\begin{figure}
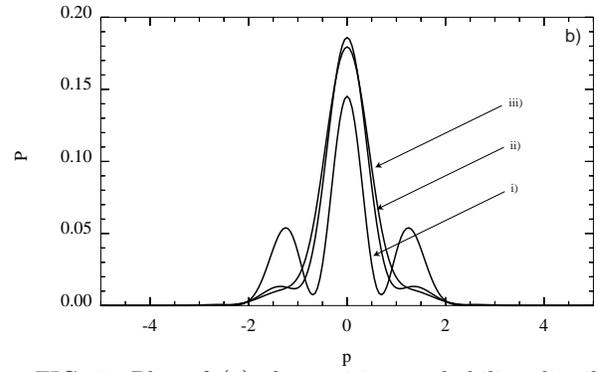

\epsfig{figure=eseriesprobx.epsf,width=.9\columnwidth}
\epsfig{figure=eseriesprobp.epsf,width=.9\columnwidth}
\narrowtext
\caption{Plot of (a) the position probability distribution $P(z)$ 
  and (b) the momentum probability distribution $P(p)$; for (i)
  $g^2=300$ with $t=0.014\tau$, (ii) $g^2=10$ with $t=0.059\tau$, (iii)
  $g^2=3$ with $t=0.19\tau$. Here $\lambda/g^2=1.5$, and the time chosen
  was informally optimised. }
\label{fig3}
\end{figure}

The results of our calculations are shown in the
Figures~\ref{fig2}-\ref{fig3}.  Figure~\ref{fig2} plots the position
($X_0$) and momentum ($X_{\pi/2}$) probability distributions versus time.
The interference fringes in the momentum probability distribution combined
with the twin peaks in the position distribution reveal the Schr\"odinger
cat state-like nature.

The formation of fringes with the evolution of the signal field from the
vacuum state is clearly evident in Figure~\ref{fig2}. As the oscillator
evolves further the fringes are washed out. The
$|i\lambda/g^2\rangle-|-i\lambda/g^2\rangle$ state, which is generated from
$|i\lambda/g^2\rangle+|-i\lambda/g^2\rangle$ with the loss of a cavity
photon, contributes more significantly as time increases, and the fringes
are lost in this case after only $0.1\tau$.

In order to establish the orders of $g$ required to obtain a clear fringe
pattern, the $P(z)$ and $P(p)$ distributions are shown in figure~\ref{fig3}
for a range of $g$ with $\lambda/g^2=1.5$. For $g^2$ greater than or of the
order of 10, interference fringes become apparent in the
transient evolution of the oscillator. The fringes (for fixed
$\lambda/g^2$) become more pronounced as $g$ increases. This is consistent
with the earlier analytical conclusions, which were based on calculations
performed in the large $g$ limit where the strength of the two-photon
nonlinearity is much greater than the single-photon cavity loss rate.

\begin{figure}
\epsfig{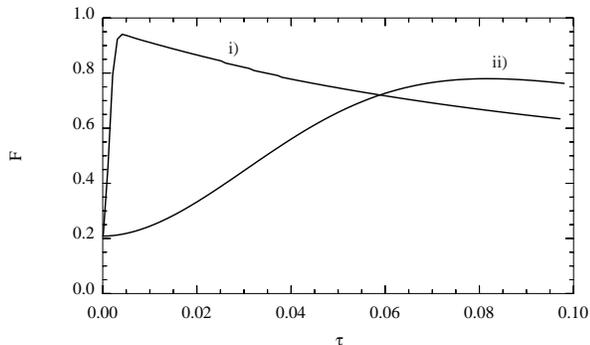}
\narrowtext
\caption{Plot of the fidelity of the generated cat state
  for (i) $g^2=300$ giving a maximum of $F\sim0.94$ and (ii) $g^2=10$
  giving a maximum of $F\sim0.78$. Here $\lambda/g^2=1.5$. The result is
  conditioned on a quadrature position measurement on the idler mode
  recording 0}
\label{fig6}
\end{figure}

As a final appraisal, we can test the fidelity of the cat state against an
ideal Schr\"odinger cat state of the form
$|i\lambda/g^2\rangle+|-i\lambda/g^2\rangle$ and this is shown in
figure~\ref{fig6} for various times. Note that it is perfectly possible to
generate a Schr\"odinger cat state not of this form but which still
constitutes a superposition of two macroscopically distinct states, hence
the fidelity in this case is an indication of the purity of the underlying
pair-coherent state rather than an indication of a good Schr\"odinger cat
state.

In this section we have established how the non degenerate parametric
oscillator operating in an adiabatically eliminated pump mode regime
will produce Schr\"odinger cat states in the transient evolution.

\section{Conclusions}

In this paper we have examined two signatures of the formation of pair
coherent states in the NDPO in the presence of linear damping. The most
direct indication is to look at the fidelity and even with large
nonlinearities we found the fidelity to be poor although higher
nonlinearity improves the fidelity. Certainly the acquired fidelity would
indicate that the generated state would be a poor candidate for the Bell
inequality test of Gilchrist {\it et. al.} even for nonlinearities as high
as given by $g^2\sim300.$

A more indirect measurement of the purity of the state is to look for the
formation of Sch\"odinger cat states which are predicted upon a conditioned
measurement on one mode given a large value for $\lambda/g^2$. The
formation of these states not only gives an indication of the presence of a
pair-coherent state but are of interest in their own right. Here we predict
that for nonlinearities characterised by $g^2\sim300$ formation of clear
Sch\"odinger cat states is possible for only $\lambda/g^2\sim1.5$. Though,
again, this parameter regime is difficult to produce experimentally.

\section{Acknowledgements}
WJM and AG both would like to acknowledge the support of the Australian
Research Council.


\end{multicols}

\end{document}